# Direct Determination of the Activation Energy for Diffusion of OH Radicals on Water Ice

A. Miyazaki[1], M. Tsuge[1], H. Hidaka[1], Y. Nakai[2], N. Watanabe[1]

[1]Institute of Low Temperature Science, Hokkaido University, Sapporo, Hokkaido 060-0819, Japan; watanabe@lowtem.hokudai.ac.jp
[2]Radiative Isotope Physics Laboratory, RIKEN Nishina Center, Saitama 351-0198, Japan

## Abstract

Using a combination of photostimulated desorption and resonance-enhanced multiphoton ionization methods, the behaviors of OH radicals on the surface of interstellar ice analog was monitored at temperatures between 54 and 80 K. The OH number density on the surface of ultraviolet (UV) -irradiated compact amorphous solid water gradually decreased at temperatures above 60 K. Analyzing the temperature dependence of OH intensities with the Arrhenius equation, the decrease can be explained by recombination of two OH radicals, which is rate-limited by thermal diffusion of OH. The activation energy for surface diffusion was experimentally determined for the first time to be 0.14 ± 0.01 eV, which is larger than or equivalent to those assumed in theoretical models. This value implies that the diffusive reaction of OH radicals starts to be activated at approximately 36 K on interstellar ice.

# 1. Introduction

Physicochemical processes like surface reactions, UV photolysis, and ion bombardment on interstellar ice are indispensable for promoting chemical evolution in the early stage of star formation. Chemical evolution on interstellar ice can be activated in a dense core of molecular clouds at temperatures as low as 10 K even without external energy inputs such as ultraviolet (UV) photons (as a review, e.g. Watanabe & Kouchi 2008; Hama & Watanabe 2013). In the first step, the hydrogenation of primordial atomic and molecular species, including $H_2$ formation, has an important role because hydrogen atoms, as light particles, can migrate and encounter the reaction partners on the dust surface even at ~10 K. Notably, molecules abundantly observed on ice, such as $H_2O$ and $CH_3OH$, have been theoretically proposed to form by the hydrogenation of oxygen and CO molecules (e.g., d'Hendecourt et al. 1985; Hasegawa et al. 1992; Cuppen & Herbst 2007; Charnley et al. 1997). Following these theoretical predictions, the formation of $H_2O$, $NH_3$, $H_2CO$, and $CH_3OH$ on the ice surface was confirmed at approximately 10 K in laboratory experiments where O, $O_2$, N, and CO reacted with H or $H_2$ (e.g., Dulieu et al. 2010; Miyauchi et al. 2008; Ioppolo et al. 2008; Hidaka et al. 2011; Watanabe & Kouchi 2002). When the temperature of ice is elevated in star-forming regions, heavier species start to move and trigger another type of reaction on the surface. Many kinds of complex organic molecules (COMs) in addition to methanol and their precursors are expected to be efficiently produced by reactions among heavier species. In particular, radical reactions are key for chemical evolution, including COMs formation, as simulated in chemical models (e.g., Hollis & Churchwell 2001). Garrod et al. (2008) explicitly demonstrated the importance of reactions of radicals such as OH, $CH_3$, and HCO for chemical complexity in hot cores and corinos. In addition, simulation experiments observed signatures for the formation of COMs through radical reactions (Butscher et al. 2017; Fedoseev et al. 2015; He et al. 2022; Ioppolo et al. 2021; Santos et al. 2022). In these experiments, the COM products were detected by infrared spectroscopy and/or temperature-programmed desorption after UV photolysis of ice mixtures or codeposition of H atoms and CO, where many radicals accumulated in/on solids. However, because of experimental difficulty in directly detecting radicals on the surface, the detailed behavior of each radical is still unknown. Since reactive radicals can cause barrierless reactions, their surface diffusion often becomes a rate-limiting process for molecular formation. Therefore, the activation energy for diffusion, $E_{diff}$, of radicals is an essential parameter for evaluating chemical evolution through radical reactions. In theoretical models, even for stable small molecules, $E_{diff}$ was often assumed to be a universal fixed fraction of desorption energy, $E_{des}$ (Cuppen et al. 2017). However, a recent experiment clearly showed that the ratios of $E_{diff}$ to $E_{des}$ are not universal but dependent on species (Furuya et al. 2022). It is not easy to quantum-chemically calculate the $E_{diff}$ for long-distance diffusion on rough amorphous surfaces relevant to realistic interstellar dust. Therefore, experimental determination of $E_{diff}$ is highly desirable.

Among various radicals, OH radicals are considered one of the most abundant radicals on ice because they can be easily produced by the surface reaction of O + H and photolysis of $H_2O$. Thus, the behaviors of OH radicals on ice would be closely related to various phenomena occurring on ice. Because the reactivity of OH radicals is very high, preparing the surface density of OH on ice enough for detection with conventional experimental methods is difficult. Therefore, methods often used for solids, such as Raman, infrared, and electron spin resonance spectroscopies, are not applicable. Although microscopic methods, such as scanning tunneling microscopy and field-emission microscopy, can detect adsorbates on the surface (e.g., Zangwill 1988; Gomer 1990; Laufon & Ho 2002), these methods are inappropriate for distinguishing between OH and $H_2O$. Furthermore, ice is not an electric conductor. The OH radicals in bulk ice were detected by near-edge, X-ray absorption, fine structure spectroscopy, but unfortunately, this method is not surface-sensitive (Laffon et al. 2006; Lacombe et al. 2006). In the present study, we apply a combination of photostimulated desorption (PSD) and resonance-enhanced multiphoton ionization (REMPI), also known as the PSD-REMPI method, which was previously developed for the detection of H (D) and $H_2$ ($D_2$) on amorphous solid water (ASW) (Kuwahata et al. 2015; Hama et al. 2012; Watanabe et al. 2010) and recently applied for detecting OH radicals on ASW (Miyazaki et al. 2020; Tsuge & Watanabe 2021; Kitajima et al. 2021). We determined $E_{diff}$ for OH radicals on compact amorphous solid water (c-ASW) by the direct detection of OH radicals.

2. Experiments

The detailed experimental setup was previously described (Miyazaki et al. 2020). The compact amorphous solid water (c-ASW) samples with approximately 100 monolayers were deposited on a sapphire disk at 100 K by introducing the vapor of freeze-pump-thaw cycled ultrapure water into an ultrahigh vacuum chamber (~$10^{-8}$ Pa). After the preparation of samples, OH radicals were produced by photolysis of $H_2O$ with UV irradiation from a conventional deuterium lamp ("H2D2" light source unit, Hamamatsu Photonics K. K.) at temperatures in the range from 54 to 80 K. The UV flux was measured by a photodiode (AXUV-100G, IRD Inc.) to be approximately 1 × $10^{13}$ photons cm$^{-2}$ s$^{-1}$ above the sample surface. The photons from the lamp photodissociate H$_2$O mainly into H + OH with minor channels, H$_2$ + O and 2H + O (Slanger & Black 1982). In addition, as secondary products, H$_2$ and O$_2$ are produced on the surface. However, these volatile photofragments and products on the surface immediately desorb when the sample temperature is well above their desorption temperatures. To avoid the possible effect of these species other than OH, the experiments were performed at temperatures from 54 K.

The OH radicals on the sample surface were detected by the PSD-REMPI method. The details of the procedure are described in Miyazaki et al. (2020). Briefly, the OH radicals on c-ASW were photodesorbed by unfocused pulsed-weak laser radiation (typically 40 μJ per pulse with an approximate 3 mm$^2$ spot on the surface) at 532 nm from a Nd:YAG laser (hereafter denoted as a PSD laser); the photon energy at 532 nm (2.33 eV) is below the dissociation energy of H$_2$O (5.17 eV). Subsequently, photodesorbed OH radicals were selectively ionized by the (2+1) REMPI process via the transition of $D^2\Sigma^-$ ($v'$ = 0) ← $X^2\Pi$ ($v''$ = 0) at approximately 1 mm above the c-ASW sample and detected by a time-of-flight mass spectrometer. The repetition rate of the laser shots was 10 Hz. Because the power density of the PSD laser shot was very weak, no heating effect on ice was observed as confirmed in the previous works (Watanabe et al. 2010; Hama et al. 2012). The detected OH intensities per shot of the PSD laser were independent of the total number of laser shots, indicating that the PSD laser shots had a minimal effect on the surface number densities of OH. In addition, to further examine the effect of PSD laser on the surface OH radicals, we performed two kinds of measurements. In the first measurement, the OH intensities were continuously monitored during UV exposure for a given period of time (the continuous PSD laser irradiation). In the other measurement, the OH intensities were measured only after the UV exposure for the same exposure time (the PSD laser irradiation only after terminating UV exposure). The intensities of OH radicals were equivalent between two measurements, showing that the PSD laser shots should have a negligible effect on the surface OH densities. Isolated neither H$_2$O nor OH absorbs a photon at 532 nm, whereas our previous work (Miyazaki et al. 2020) revealed that OH-(H$_2$O)$_n$ complex having three hydrogen bonds with neighboring H$_2$O molecules can absorb a photon at approximately 532 nm, leading to photodesorption. According to quantum chemical calculations, the binding energies of OH on the ASW surface range from 0.06 to 0.74 eV depending on the number of dangling-H or dangling-O atoms on the binding site (Miyazaki et al. 2020). Notably, the OH radicals adsorbing to the surface through three hydrogen bonds should have nearly the strongest binding energy. In the present experiments, using PSD laser radiation at 532 nm, we can selectively monitor OH radicals trapped in deep binding sites, and the detected OH intensities, $I_{OH}$, are proportional to the surface densities, [OH], of these deeply bound OH radicals. It should be noted that the PSD of OH radical is caused by one-photon absorption process but not by thermal process.

## 3. Results and Discussion

Figure 1 shows the (2+1) REMPI spectrum for OH desorbed from c-ASW during UV irradiation at 70 K. The observed spectrum was well reproduced with rotational temperatures of OH in the range of 120 to 200 K by using the PGOPHER program (Western 2017) regardless of the c-ASW temperature, indicating that the rotational distribution is determined by photodesorption dynamics rather than the sample temperature. Although determination of the mechanism is ambiguous, this kind of independence of rotational temperature has also been reported in experiments on $H_2O$ photodesorption from ice (Hama et al. 2016). Hereafter, the OH intensities are represented by the area of the strongest peak at 244.164 nm. The OH intensities immediately increased upon turning the UV lamp on and reached a steady state after approximately several minutes of irradiation. The OH intensities did not change during the measurements, typically greater than 1 hour of UV irradiation. We monitored the OH intensities on the surface of c-ASW at temperatures between 54 and 80 K in the steady state during UV irradiation, as shown in Figure 2(a). The OH intensities gradually decrease with the temperature of c-ASW from approximately 60 K to 80 K. This result was conserved regardless of procedure of heating or cooling the sample. Reasonably assuming that the surface composition is dominated by $H_2O$ and OH only (Miyazaki et al. 2020), the OH surface density, [OH], in deeply bound sites under steady state conditions at a given temperature can be expressed by

$$0 = \frac{d[OH]}{dt} = f\sigma c[H_2O] - 2k_{OH\text{-}OH}[OH]^2 - k_{des}[OH], \qquad (1)$$

where $f$, $\sigma$, c, $k_{OH\text{-}OH}$, and $k_{des}$ are the UV flux, dissociation cross-section of $H_2O$ to H + OH, a factor for remaining OH on the surface at photodissociation, rate constant of OH–OH recombination, and desorption rate of OH radical, respectively. Some fraction of products from recombination would desorb by so-called chemical desorption (Williams 1968; Garrod et al. 2007). Because the dissociation cross-section and the factor "c" should be independent from the sample temperature, the decrease in OH intensities with temperature in Figure 2(a) can be attributed to recombination or desorption processes, whose rate strongly depends on temperature. To evaluate the dominant process for the decrease, we analyze the data with both recombination- and desorption-dominating cases. In the former case, the third term of the right side in Equation (1) is set to zero. Furthermore, because the recombination reaction is generally barrierless and its rate is diffusion-limited, $k_{\text{OH-OH}}$ can be written by

$$k_{\text{OH-OH}} = D_{\text{s}} = D_0 \exp\left(-\frac{E_{\text{diff}}}{k_{\text{B}}T}\right), \quad (2)$$

where $D_0$ and $k_{\text{B}}$ are the prefactor of the diffusion coefficient and the Boltzmann constant, respectively. As a result, Equation (1) can be rewritten as

$$-\ln[OH]^2 = \ln D_0 - \ln\frac{\alpha}{2} - \frac{E_{\text{diff}}}{k_{\text{B}}T}, \quad (3)$$

where $\alpha = f\sigma[H_2O]$. Since $[OH] \propto I_{\text{OH}}$, the OH intensities can be replotted with $-\ln(I_{\text{OH}}^2)$ as a function of $T^{-1}$ (Figure 2(b)), and $E_{\text{diff}}$ is derived to be $0.14 \pm 0.01$ eV ($1650 \pm 60$ K) by linearly fitting the decreases at temperatures between 64 and 76 K to Equation (3). Recently, the average binding energy of OH radicals for 18 binding sites on ASW was estimated to be 0.37 eV by quantum chemical calculations on the assumption that 18 kinds of sites are equally distributed on the surface (Miyazaki et al. 2020). Simply taking an $E_{\text{des}}$ of 0.37 eV, our obtained value for $E_{\text{diff}}$ is approximately $0.35 E_{\text{des}}$. However, note that our obtained value represents the activation energy to overcome the diffusion-limiting deep binding sites. In the desorption dominating case, the second term on the right side of Equation (1) is zero. By fitting the Arrhenius plots of $-\ln(I_{\text{OH}})$ as a function of reciprocal temperature, $E_{\text{des}}$ can be determined to be $0.07 \pm 0.01$ eV (refer to Appendix 1), which is too small compared with the prediction from the calculated value of 0.37 eV for $E_{\text{des}}$. Moreover, when using 0.07 eV as the activation energy of thermal desorption, the OH radicals cannot remain on the surface at 50 K in the experimental duration when the prefactor of thermal desorption is $\sim 10^{12}$ s$^{-1}$. Consequently, we conclude that the decrease in OH intensity is caused by OH–OH recombination and that the activation energy for surface diffusion is determined to be $E_{\text{diff}} = 0.14 \pm 0.01$ eV ($1650 \pm 60$ K).

The activation energy for diffusion is also derived by another experiment to cross-check the derived value above. The c-ASW samples were irradiated with UV for 30 min at temperatures from 66 to 76 K in increments of 2 K. After the termination of UV, the attenuation of OH intensities was measured at each temperature, as shown in Figure 3(a). Assuming the recombination dominated case, the variation in OH surface densities after terminating the supply of OH can be expressed by

$$\frac{d[OH]}{dt} = -2k_{\text{OH-OH}}[OH]^2. \quad (4)$$

Integration of Equation (4) gives the following equation:

$$[\text{OH}] = \frac{1}{2k_{\text{OH-OH}}t + [\text{OH}]_0^{-1}}. \tag{5}$$

The value of $k_{\text{OH-OH}}$ can be determined by fitting the $I_{\text{OH}}$ data points to Equation (5). Furthermore, the values of $k_{\text{OH-OH}}$ at each temperature are shown in the Arrhenius plot (Figure 3(b)). From the slope of the linear fitting, the activation energy for OH diffusion is determined to be 0.13 ± 0.01 eV (1540 ± 80 K), which is consistent with the previously obtained value. In the derivation under the previously described steady state, the photodesorption of OH by UV irradiation was implicitly disregarded in Equation (1). The consistency in the results between two different experiments indicates the validity of this treatment at temperatures above 64 K. In addition, when the UV-photolyzed ice was capped by approximately one monolayer of water molecules, no OH was detected, indicating the detected OH was not coming from inside the ice. In Appendix 3, we also analyzed the data on the assumption of desorption dominating case. It does not work as well as the above experiment.

We confirmed the occurrence of recombination during the OH decrease by a different experiment. OH + OH recombination on ice should produce $H_2O_2$ (Yabushita et al. 2008) with $H_2O$ + O (Redondo et al. 2020). Therefore, in our scenario, at temperatures above approximately 65 K, the number density of $H_2O_2$ should increase. Unfortunately, the REMPI method cannot be applied for the detection of $H_2O_2$ because no appropriate intermediate state exists. Therefore, photodissociation of $H_2O_2$ to produce 2OH was utilized to monitor $H_2O_2$. According to the calculated absorption cross-section of $H_2O_2$ at the air–water interface, the photodissociation of $H_2O_2$ into 2OH on water ice occurs at wavelengths shorter than 300 nm and is most efficient at 200 nm with the cross section of ~8 × 10$^{-19}$ cm$^2$ (Ruiz-López et al. 2021). That is, photon at 532 nm cannot dissociate $H_2O_2$. Replacing the radiation at 532 nm (Nd:YAG) with laser radiation at 200 nm from an OPO laser as the PSD laser, $H_2O_2$ can be photolyzed into 2OH. Some fraction of OH fragments should desorb upon photolysis and be subsequently detected by the REMPI method. We confirmed that using the 200-nm PSD laser, OH radicals were detectable from $H_2O_2$ solid, which was produced by a background deposition of $H_2O_2$ vapor sublimated from a urea hydrogen peroxide sample kept at room temperature (Pettersson et al. 1997). In this experiment, the c-ASW samples were irradiated with the UV lamp for 1 min at 55 K to produce a certain amount of OH on the surface. Next, the PSD (200 nm)-REMPI method was applied to continuously detect OH photofragments from the surface during raising the temperature at a ramping rate of 5 K min$^{-1}$. The duration of UV irradiation was set as short as possible to suppress the yields of $H_2O_2$ during UV photolysis, which obscures the variation of $H_2O_2$ surface density due to the OH–OH recombination reaction at warming up. The wavelength of the REMPI laser was set to the same wavelength utilized in the first experiment. Figure 4 shows the variation in the OH intensity originating from $H_2O_2$ photodissociation with replots from Figure 2(a) as functions of the substrate temperature. Note that photons at 200 nm cannot dissociate $H_2O$ molecules, that is, without the UV process, no OH was observed solely by irradiation at 200 nm. As shown in Figure 4, the intensities of OH fragments from $H_2O_2$ increase above 65 K, while those of OH detected by 532 nm PSD-REMPI decrease. This result supports the enhancement of OH + OH recombination to form $H_2O_2$ in this temperature region.

As previously discussed in H-atom diffusion experiments (Hama et al. 2012), there are various binding energies of OH adsorption sites on c-ASW. Moreover, quantum chemical calculations showed a wide range of binding energies of radicals (OH, HCO, and $CH_3$) for various adsorption sites on ice surfaces (Sameera et al. 2017, 2021; Miyazaki et al. 2020). As previously described, PSD-REMPI at 532 nm detects OH radicals strongly bound to the surface. Therefore, our determined activation energy represents the diffusion rate-limiting value under the present experimental conditions. There should exist various shallower adsorption sites where OH can move at lower temperatures.

We briefly discuss the temperature independent plateau of OH intensities at temperatures below about 60 K in Figure 2(a). In the present experiments, using the value of UV flux and the cross section of $10^{-18}$ cm$^2$ for OH production by photolysis of $H_2O$ at around Lyman α (Lee et al. 1987), the OH coverage after about 1 h of UV exposure can be estimated to be about 0.05 if all OH can remain intact. Actually, it was reported that photodesorption rates of OH radicals are comparable to the OH photoproduction rates from $H_2O$ in the UV region (Cruz-Diaz et al. 2018), implying that the steady-state coverage would be rather smaller than 0.05. Most importantly, we found that our detected OH yields over several minute exposure reached a similar level within a factor of 2 to those over 1 h at the steady state. This quick saturation of OH yields was also seen in the experiments by Cruz-Diaz et al. (2018). These findings indicate that the surface number density of OH is settled in the steady state even at low coverage of OH by the balance between photoproduction and loss by direct UV photodesorption of OH adsorbates and recombination such as OH + OH → $H_2O_2$ (Yabushita et al. 2008) or $H_2O$ + O (Redondo et al. 2020). $H_2O_2$ can be a source of OH reproduction by UV photolysis for which the photodissociation cross section is around $5 \times 10^{-18}$ cm$^2$ for the photons from the UV lamp (Suto and Lee 1983). These products may desorb at the formation reaction (Yabushita et al. 2008). Although we do not exclude the OH–OH recombination even below 60 K through the rapid diffusion of energetic OH fragment, it is unlikely to significantly contribute to OH loss. We deduce that direct photodesorption of OH radical by UV photons would be the main route for OH loss below 60 K. Unfortunately, there is no report on the OH photodesorption from ice in UV region. It is very difficult to measure it experimentally because of competition with OH desorption originated from photodissociation of $H_2O$ and $H_2O_2$. To evaluate the constant intensities below 60 K, equation (1) is too simplified. However, because there are many unknown factors, discussion on that is beyond the scope of the present study. Nevertheless, the equation (1) works very well to explain the decrease of OH at temperatures above 60 K.

The activation energy of OH radical diffusion is meaningful for evaluating reactions with less mobile species than OH, where the diffusion of OH controls the reaction rates. Previous chemical models assumed activation energies for OH diffusion as taking different $E_{des}$ and factors, β, in $βE_{des}$ (Table 1). For example, Ruffle & Herbst (2000) used activation energies for the diffusion of 378 or 970 K with $E_{des}$ = 1260 K and β = 0.3 or 0.77, respectively. Garrod et al. (2008) employed $E_{diff}$ of 1425 K as $0.5E_{des}$. The $E_{des}$ values of 4600 K (Wakelam et al. 2017) and 5698 K (Minissale et al. 2022) were also adopted in the literature. If we assume $E_{diff} = 0.35E_{des}$ (Garrod et al. 2017), these values give $E_{diff}$ = 1610 and 1994 K, respectively. The $E_{diff}$ values assumed in theoretical models have a wide range depending on the adopted $E_{des}$ value and β factor. As mentioned earlier, Furuya et al. (2022) suggested that the β factor is molecule dependent, and therefore the experimental determination of $E_{diff}$ for OH radical is critically important for chemical models to precisely describe the bahavior of OH radicals on ice.

4. Astrophysical Implication

Simply assuming that the activation energy takes the single value of our derived 1650 K, a temperature window where the OH reaction can be activated is proposed. The factor $D_0$ in Equation (2) can be written by $D_0 = a^2ν$, where a and ν are the distance of single hopping and a frequency factor. The "a" value would be approximately 0.3 nm, which is the average distance between two $H_2O$ molecules on ice. Although the precise determination of ν is not easy, it is simply set to $ν = 10^{12}$ s$^{-1}$. Figure 5 shows the surface temperature dependence of the diffusive area of OH at each $E_{diff}$ value with the time required for 100 nm diffusion, which is equivalent to the characteristic diameter of dust. When $E_{diff}$ = 1650 ± 60 K, the OH radicals can migrate over 100 nm on the dust surface within $10^5$ yrs at temperatures above 36 ± 1 K. That is, the present results suggest that OH migration starts at 36K OH chemistry on ice becomes important especially in the warming-up phase of star-forming regions. Note that we do not exclude the possibility that the OH radicals move to some extent through shallower sites even at temperatures below 36 K. Therefore, the diffusive areas described in Figure 5 should be the lower limits for realistic interstellar ice.

## 5. Summary

Employing the PSD-REMPI method, we monitored OH radicals strongly bound to water ice surface at temperatures from 54 to 80 K. The surface densities of OH radicals reached at a steady state condition during continuous UV irradiation due to the balance between the OH production and the loss. It was found that the OH intensities decreased at temperatures above 60 K, where the diffusive OH-OH recombination becomes prominent. From the temperature dependence of steady-state OH radical intensities under UV irradiation and the temporal decay of OH radical intensities at fixed temperatures after UV irradiation, we determined for the first time the activation energy of OH diffusion on ice to be $E_{\text{diff}} = 0.14 \pm 0.01$ eV (1650 ± 60 K). Although the activation energies should have a distribution, this value represents the diffusion barrier for strongly bound OH radicals, which limits the rate of diffusive reactions. Taking the $E_{\text{diff}}$ value determined in this work, we suggest that the OH migration would be activated at around 36 K. This work demonstrated that the PSD-REMPI method is quite powerful in studying behavior of radicals on the water ice surface.

Table 1. Comparison of the $E_{\text{diff}}$ values of OH radicals on ASW used in theoretical models and determined in this study.

|  | $E_{\text{diff}}$ | $E_{\text{des}}$ | method |
|---|---|---|---|
| Ruffle & Herbst (2000) | 0.03 eV (378 K) | 0.11 eV (1260 K) | $0.3 E_{\text{des}}$ |
| Ruffle & Herbst (2000) | 0.08 eV (970 K) | 0.11 eV (1260 K) | $0.77 E_{\text{des}}$ |
| Garrod et al. (2008) | 0.12 eV (1425 K) | 0.25 eV (2850 K) | $0.5 E_{\text{des}}$ |
| Wakelam et al. (2017) | [0.14 eV (1610 K)][a] | 0.40 eV (4600 K) | $0.35 E_{\text{des}}$[a] |
| Minissale et al. (2022) | [0.17 eV (1994 K)][a] | 0.49 eV (5698 K) | $0.35 E_{\text{des}}$[a] |
| This study | 0.14 eV (1650 K) | ----- | experiment |

[a] $E_{\text{diff}}$ values were not derived in these studies and, therefore, we adopted $E_{\text{diff}} = 0.35 E_{\text{des}}$ (Garrod et al. 2017) to calculate $E_{\text{diff}}$.

This work was supported in part by JSPS KAKENHI grant number 22H00159. AM is grateful for the support from JST SPRING, Grant Number JPMJSP2119.

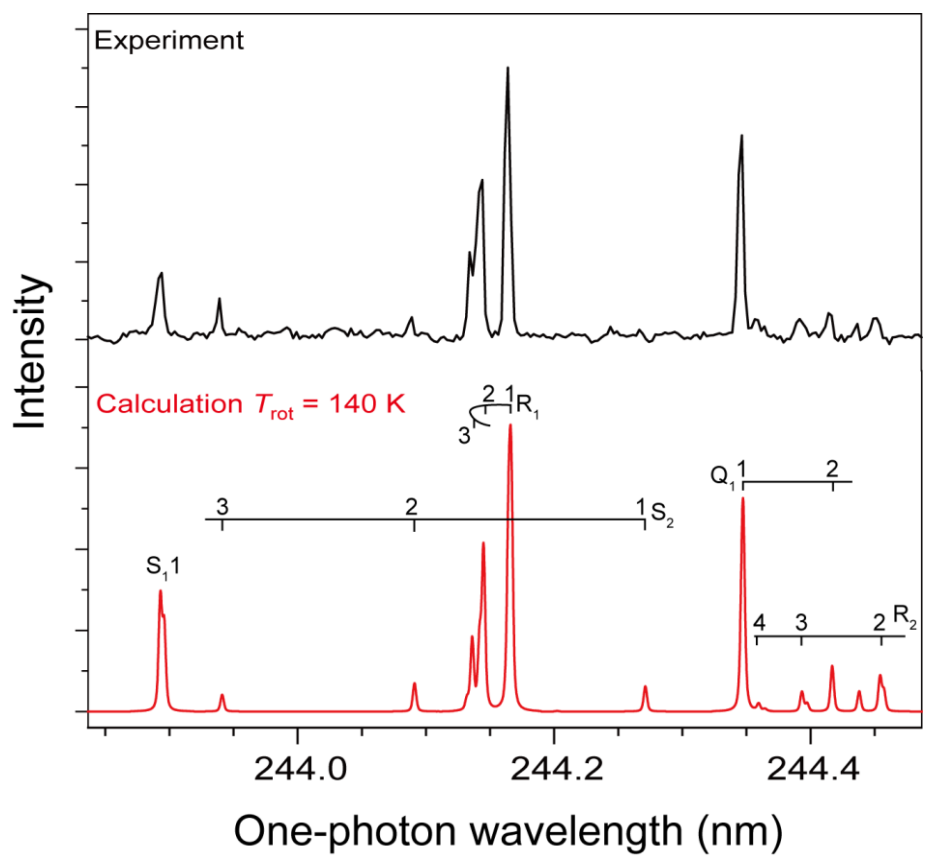

**Figure 1.** PSD-(2+1)REMPI spectrum (upper) obtained from c-ASW during UV irradiation at 54 K and the simulation calculation (lower) for OH in the transition of $D^2\Sigma^-$ ($v' = 0$) ← $X^2\Pi$ ($v'' = 0$) with a rotational temperature of 140 K. Adopted from Tsuge & Watanabe 2021 and Miyazaki et al. 2020. Although the spectrum at 140 K best reproduces the experiment, those at 120 and 200 K also achieve fairly good fits (see Fig 1 of Miyazaki et al. 2020). S, R, and Q on the simulation spectrum denote spectral branches.

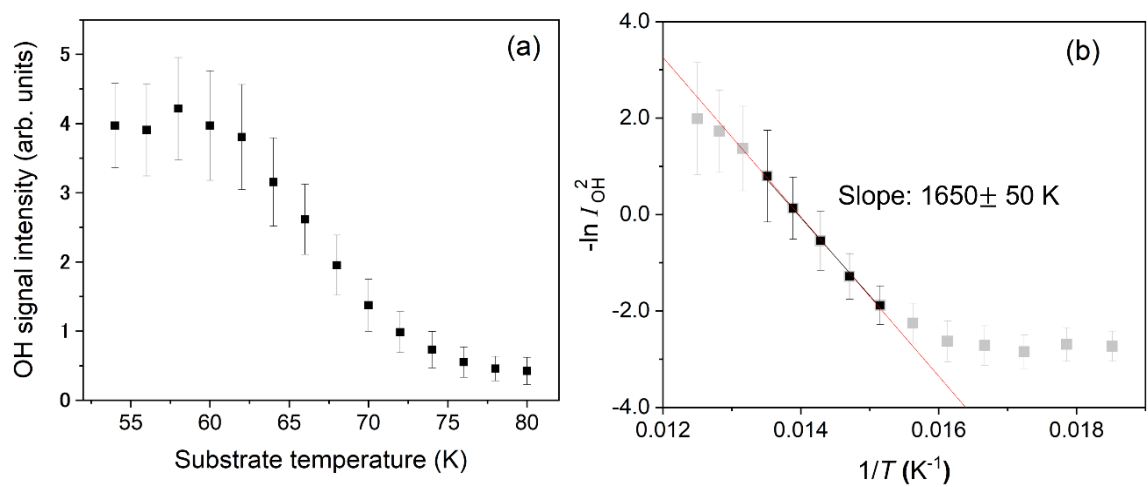

**Figure 2.** (a) PSD-(2+1)REMPI intensities for OH detected from c-ASW during UV irradiation in the steady state at each temperature. (b) Arrhenius plots in the square of OH intensities on the assumption that the OH intensities are dominated by diffusive recombination of OH radicals (Equation (3)). The red line represents the result of linear fitting with data points at temperatures of 66, 68, 70, 72, and 74 K, which are represented by black symbols.

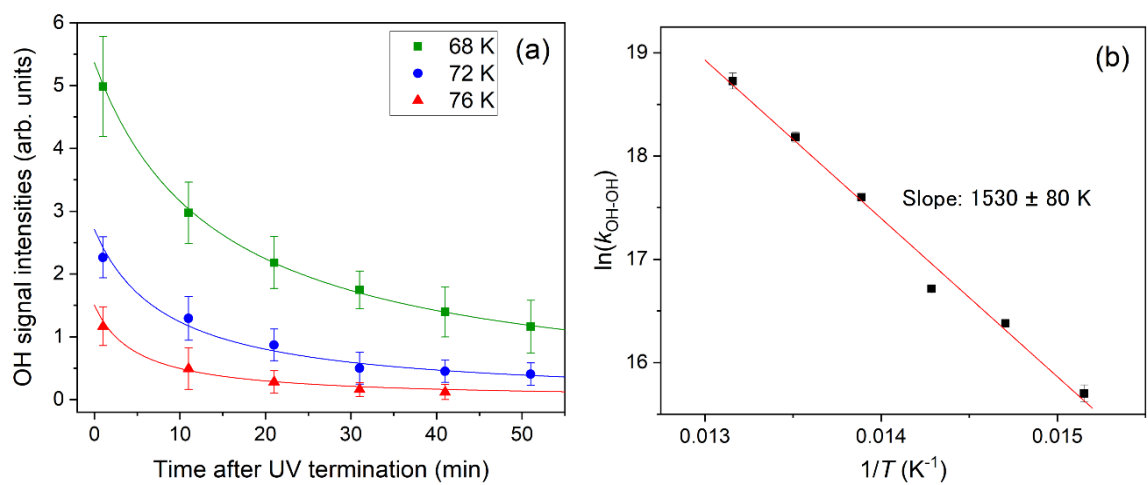

**Figure 3.** (a) Attenuation of OH intensities after irradiation of c-ASW for 30 min at temperatures of 68, 72, and 76 K. To avoid complexity in the figure, the data for other temperatures are shown in Figure A2. The solid lines represent the fitting results with Equation (4). (b) $k_{OH-OH}$ values derived from the fitting results shown by solid lines in (a) were plotted as a function of $1/T$. The red line denotes the linear fitting results.

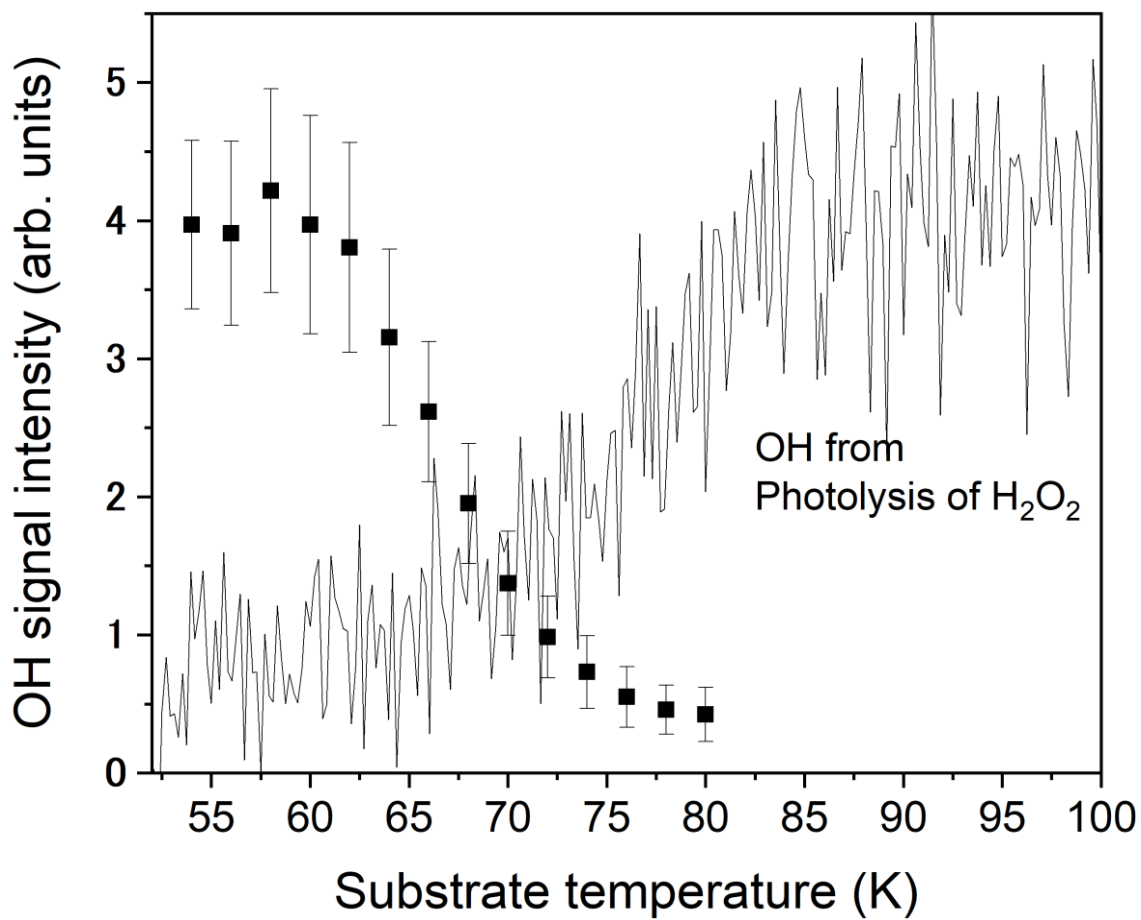

Figure 4. The solid squares represent the steady-state OH intensities during UV exposure obtained with the PSD (532 nm)-REMPI method (replots from Figure 2(a)). The solid line shows the PSD (200 nm)-REMPI spectrum for OH originating from $H_2O_2$ photolysis at 200 nm. The intensities in the solid line are normalized to compare with the data points denoted as solid squares.

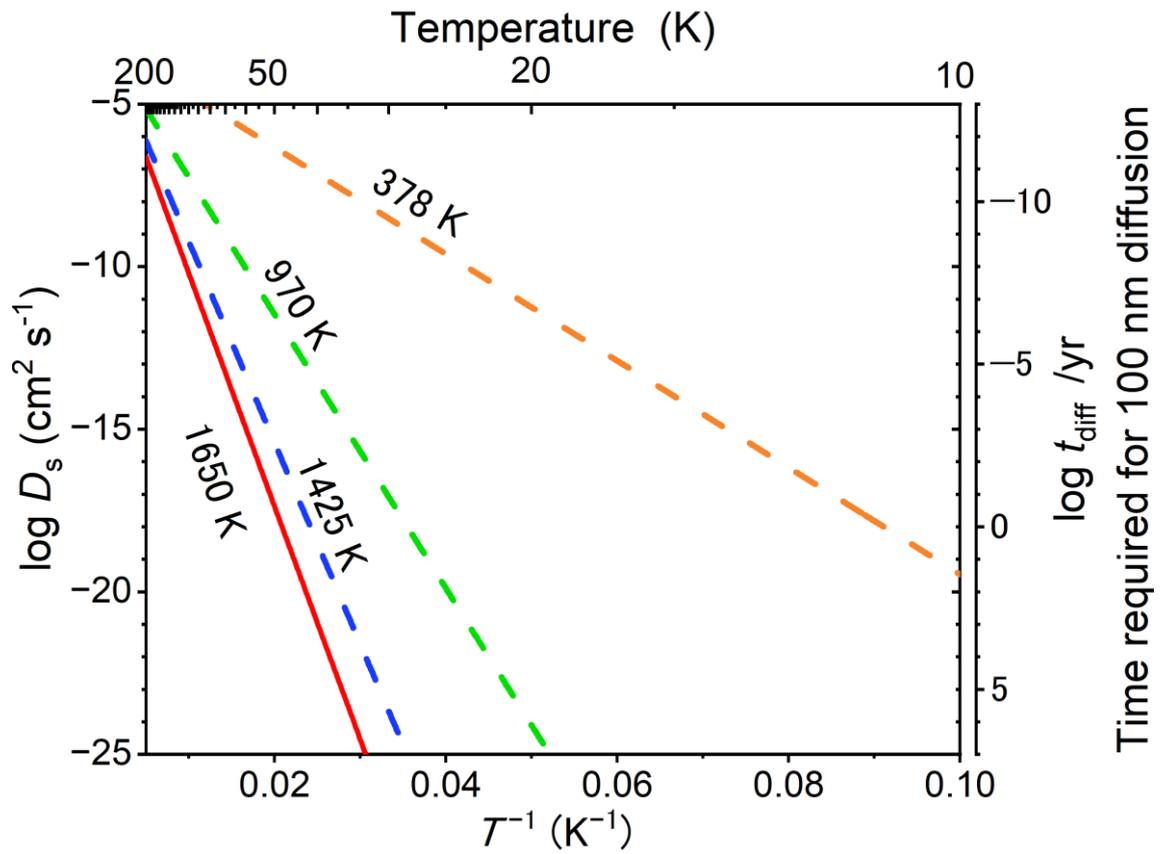

Figure 5. Diffusive area ($D_s$, Equation (2)) of OH radicals on interstellar ice with different activation energies at various temperatures. The right vertical axis represents the time required for diffusion over the typical dust size of 100 nm ($t_{diff}$).

## Appendix

### 1. Data analysis for the desorption-dominated case in steady-state experiments

Putting the second term of the right side in Equation (1),

$$0 = \frac{d[OH]}{dt} = f\sigma c[H_2O] - k_{des}[OH], \quad (6)$$

As $k_{des} = \nu_{des}\exp\left(-\frac{E_{des}}{k_B T}\right)$, where $\nu_{des}$ is the prefactor of desorption, the equation can be rewritten as

$$-\ln[OH] = \ln \nu_{des} - \ln \alpha - \frac{E_{des}}{k_B T} \quad (7).$$

When the OH intensities are replotted with $-\ln(I_{OH})$, Figure A1 is obtained. By linear fitting, $E_{des}$ is derived to be 0.07 ± 0.01 eV. Because [OH] is proportional to $I_{OH}$, the -ln($I_{OH}$) values are shifted from ln[OH] by a constant value. Therefore, $E_{des}$ can be derived from the plot of -ln($I_{OH}$) versus 1/T.

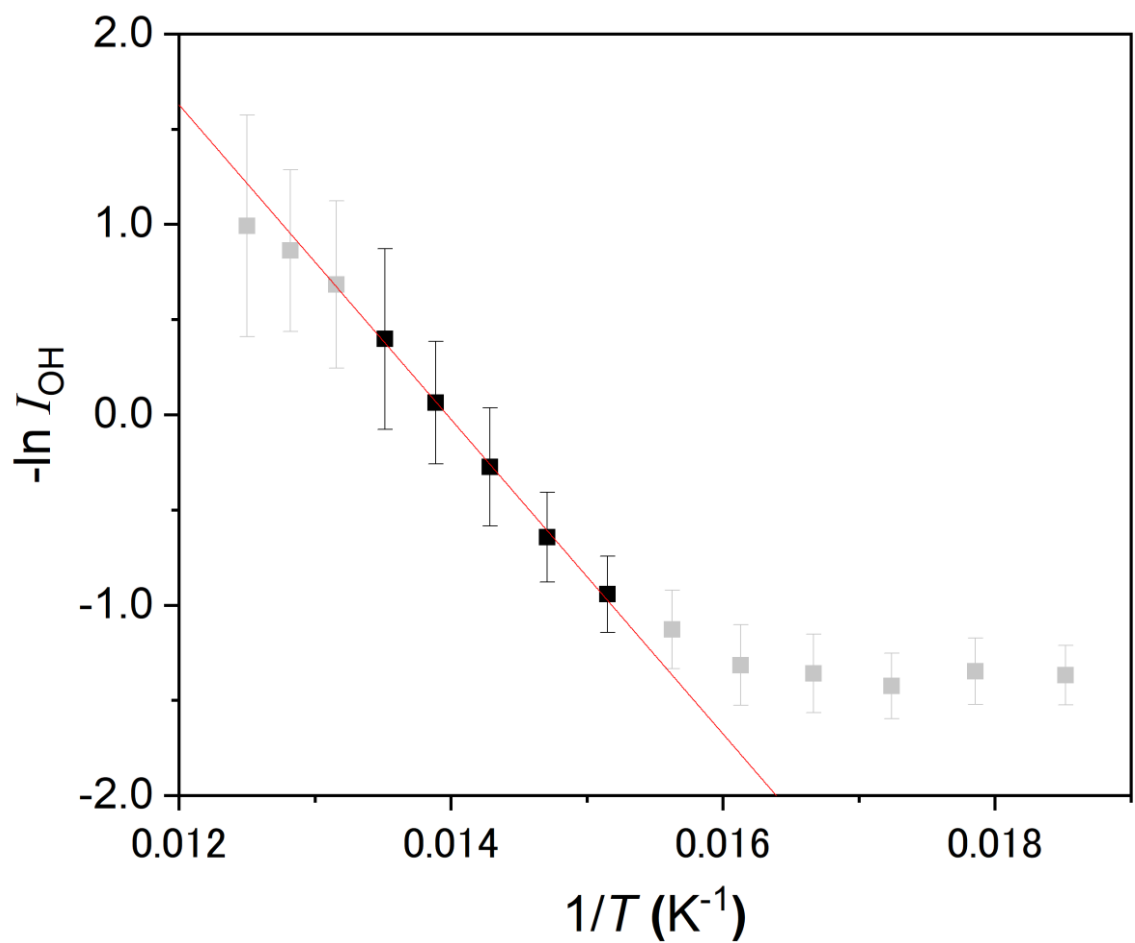

**Figure A1.** Arrhenius plots of the OH intensities with the assumption that the OH intensities are dominated by desorption of OH radicals (Equation (7)). The red line represents the linear fitting results with data points at temperatures of 66, 68, 70, 72 and 74 K, which are represented by black symbols.

2. The data for the attenuation of OH intensities after 1 min of UV irradiation are not shown in Figure 3 (a)

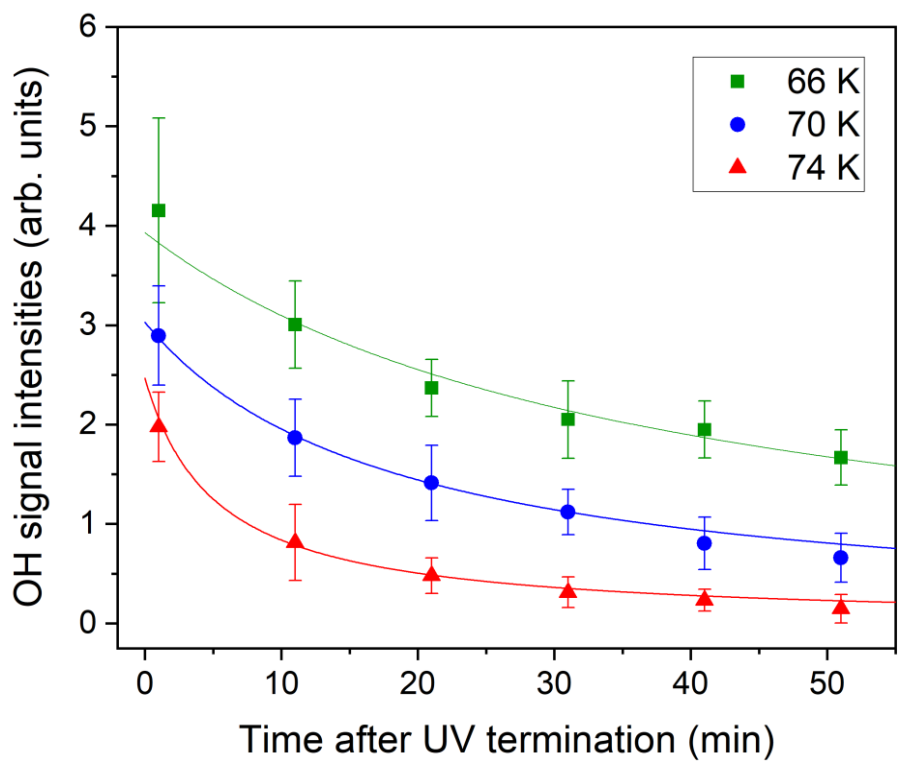

**Figure A2.** Data for the attenuation of OH intensities after 1 min of UV irradiation at temperatures of 66, 70, and 74 K.

## 3. Analysis of data in Figure 3 (a) with desorption-dominated case

Assuming the attenuation of OH intensities is dominated by OH desorption, the data in Figure 3 are fitted to the equation,

$$\frac{dI_{\text{OH}}}{dt} = -k_{\text{des}}I_{\text{OH}} \quad (8).$$

As seen Figure A3, the fittings to equation (8) were worse than those to equation (5). Nevertheless, we derived the activation energy for the desorption to be 0.07 eV, which is consistent with the result of the first experiment. Again, this value is too small for the desorption. Therefore, we conclude that the attenuation is dominated by recombination.

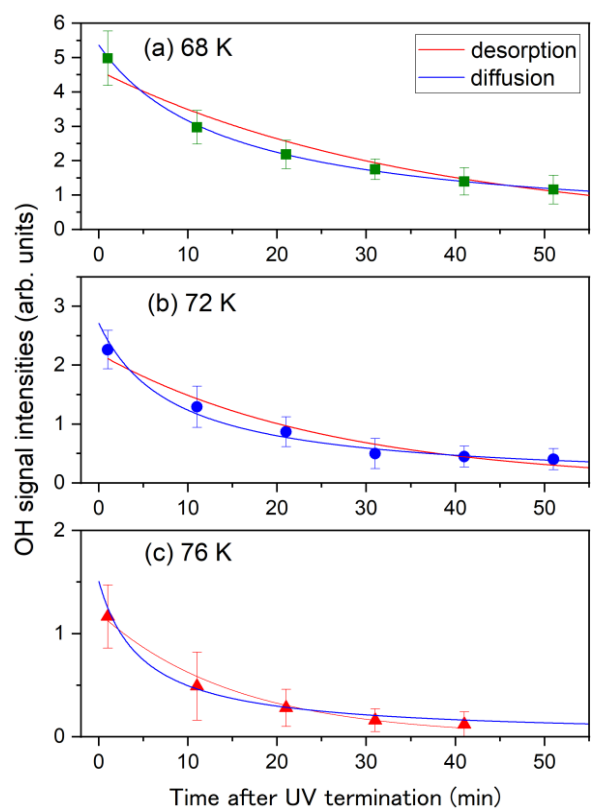

Figure A3. Fitting results for data in Figure 3 (a). Red lines represent the fitting to equation (8). Blue lines are fitting to equation (5).